\documentclass[10pt,aps,prl,twocolumn,superscriptaddress,nobibnotes,nodoi,reprint,longbibliography]{revtex4-2}

\usepackage{graphicx}
\usepackage{amsmath,physics,bm,float}
\usepackage{xcolor}
\usepackage{soul}
\usepackage[colorlinks=true,citecolor=blue,urlcolor=blue,linkcolor=blue]{hyperref}
\usepackage{amssymb}
\usepackage{natbib}
\usepackage{upgreek}
\usepackage{mathtools}
\usepackage{siunitx}
\usepackage[normalem]{ulem}

\DeclareGraphicsExtensions{.pdf,.png,.jpg}
\graphicspath{{./Pictures/}}

\newcommand{\hhu}{Institut f\"ur Theoretische Physik II: Weiche Materie, Heinrich-Heine-Universit\"at D\"usseldorf, Universit\"atsstra{\ss}e 1, D-40225 D\"usseldorf,
	Germany}
\newcommand{\tud}{Institut f\"ur Physik der kondensierten Materie, Technische Universit\"at Darmstadt, Hochschulstra{\ss}e 8, D-64289 Darmstadt, Germany}

\newcommand{\sau}{Theoretische Physik, Universit\"at des Saarlandes, Campus E26, D-66123 Saarbr\"ucken, Germany
}
\newcommand{\qute}{Center for Quantum Technologies (QuTe), Saarland University, Campus, 66123 Saarbr\"ucken, Germany}
\newcommand{\ukl}{Department of Physics and Research Center OPTIMAS, RPTU University Kaiserslautern-Landau, Erwin-Schr\"odinger-Stra{\ss}e 46, D-67663 Kaiserslautern, Germany}
\newcommand{\jug}{Institut f\"ur Physik, Johannes Gutenberg-Universit\"at Mainz Staudingerweg 9, D-55128, Mainz, Germany}
 
\begin{document}
\title{Quantization of the classical Mpemba effect}

\author{Jannis Melles}
\affiliation{\hhu}

\author{Hartmut L\"owen}
\affiliation{\hhu}

\author{Benno Liebchen}
\affiliation{\tud}

\author{Michael te Vrugt}
\affiliation{\jug}

\author{Giovanna Morigi}
\affiliation{\sau}
\affiliation{\qute}

\author{Artur Widera}
\affiliation{\ukl}

\author{Alexander P.\ Antonov}
\email{alexander.antonov@hhu.de}
\affiliation{\hhu}

\date{\today}
\begin{abstract}
	   The Mpemba effect refers to the counterintuitive phenomenon that an initially hot system can freeze faster than an initially warm one. Recent years have brought major advances in both classical and quantum realizations, with the asymmetric bistable potential emerging as the paradigmatic classical benchmark because its mechanism is especially transparent and controllable. Yet for precisely this benchmark problem, the impact of quantization remains unexplored. Here we show that quantization shifts Mpemba behavior to qualitatively new regimes, moving it to ultra-cold temperatures that are orders of magnitude lower than those relevant for classical thermal barrier crossing. In addition, quantization produces inverse and double inverse Mpemba effects that are absent in the corresponding classical dynamics.
Our results establish quantization as a robust route to quantum-enabled Mpemba effects inaccessible in classical regimes.
\end{abstract}

\maketitle

A counterintuitive relaxation phenomenon in which an initially hotter system relaxes faster toward a colder environment than the same system prepared at intermediate temperatures is widely referred to as the Mpemba effect \cite{mpemba1969cool, Teza/etal:2026}. While this was once viewed largely as a curiosity, it is now recognized as a genuine nonequilibrium effect. It has by now been identified in a broad range of classical and quantum systems, including granular media \cite{Lasanta/etal:2017, Megias/Santos:2022, Biswas/etal:2023, Antonov/Lowen:2026}, colloids \cite{lu2017nonequilibrium, kumar2020exponentially, Kumar/etal:2022, malhotra2024double}, quantum gases strongly interacting with light \cite{Keller/Morigi:2018} and spin models \cite{Carollo/etal:2021, Chatterjee/etal:2023, Nava/Egger:2024, Joshi/etal:2024, Moroder/etal:2024, Xu/etal:2025, schnepper2025experimental, li2025quantum, Bao/Hou:2025, Blom/etal:2026}.\ Within the classical literature, the asymmetric bistable-potential setup theoretically studied by Lu and Raz \cite{lu2017nonequilibrium} and experimentally realized by Kumar and Bechhoefer \cite{kumar2020exponentially} has emerged as a particularly important benchmark. There, the effect is both physically transparent and experimentally controllable: relaxation is governed by thermally activated transfer between the two wells, so the anomalous relaxation can be related directly to the population imbalance between the two basins \cite{lu2017nonequilibrium, kumar2020exponentially, Kumar/etal:2022, malhotra2024double}. It is precisely this combination of simplicity and control that has made the bistable potential a paradigmatic platform for the classical Mpemba effect \cite{Lasanta/etal:2017, Biswas/etal:2023, lu2017nonequilibrium, hayakawa2026mpemba}.

\begin{figure}[htp!]
    \centering \includegraphics[width=0.9\columnwidth]{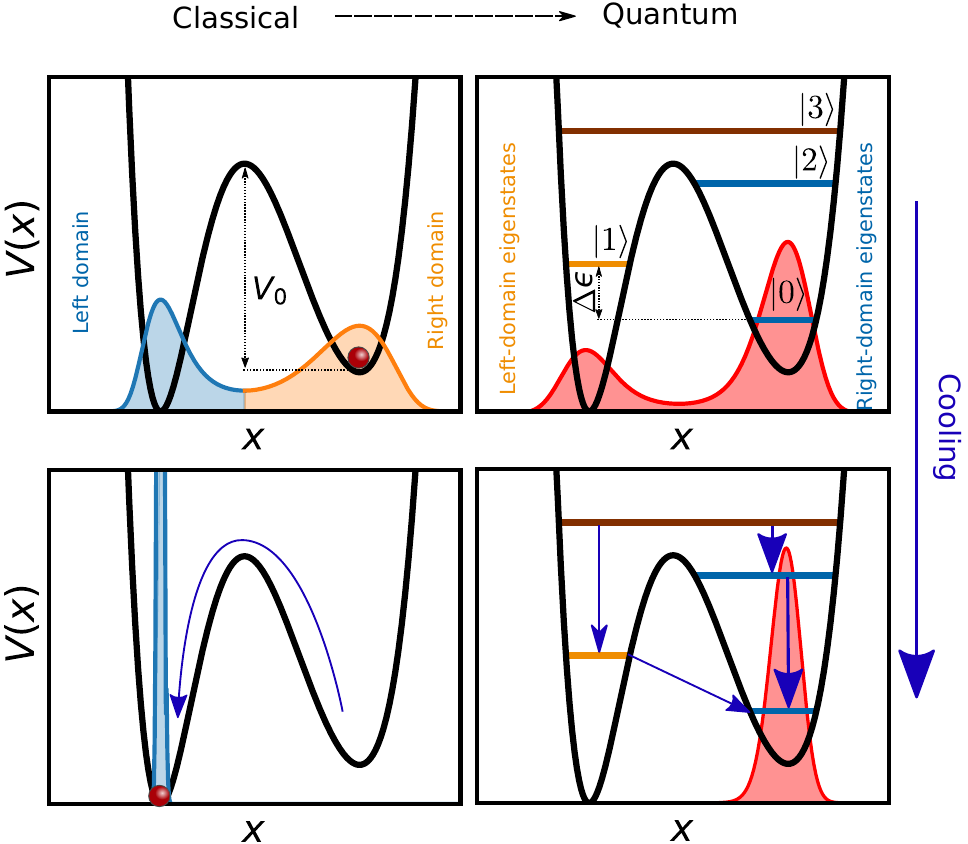} 
     \caption{Schematic representation of the cooling process in a double well potential. The classical and quantum processes are illustrated on the left and right panels, respectively. For a classical particle, escape from a metastable state (right domain) to a global one (left domain) follows the Arrhenius law and generally requires thermal energies comparable to barrier height $V_0$ to be achieved in a finite time.
     In the quantum regime, tunneling between bound quantum states enables transitions between the wells at energies $\Delta\epsilon$ well below the barrier height, giving rise to additional relaxation channels. Fast and slow transitions in the lower panel are depicted by one-way arrows with a big and small point, respectively.
    }
    \label{fig:1}
\end{figure}

Quantum studies of the Mpemba effect, however, have followed a very different route. To date, most work has focused on systems with entirely different microscopic structure, especially spin models and open quantum systems analyzed through Liouvillian relaxation modes \cite{Carollo/etal:2021, Chatterjee/etal:2023, Nava/Egger:2024, Joshi/etal:2024, Moroder/etal:2024, Xu/etal:2025, schnepper2025experimental, li2025quantum, Bao/Hou:2025, yu2025quantum, Summer/etal:2026, Ares/etal:2025, alyuruk2025thermodynamic}. This has left an important gap in our understanding: The Mpemba effect is well understood classically in the paradigmatic bistable setup -- which makes the Mpemba effect intuitively accessible -- but we still do not know what quantization does to that benchmark problem itself. Does quantization suppress anomalous relaxation, or favor it? Might it even induce Mpemba behavior in regimes where no classical effect is present?

Here we address these questions by studying the paradigmatic double-well Mpemba setup across the classical and quantum regimes within the same underlying Hamiltonian. Our results show that the Mpemba effect in the paradigmatic double well setup does not only survive quantization, but strikingly, it changes qualitatively, and enters entirely new regimes. In particular, the quantum Mpemba effect appears on temperature scales orders of magnitude below those relevant classically. Beyond this, the quantum system displays anomalous heating without a classical counterpart at the same target temperature, including an inverse Mpemba effect and even a double-inverse Mpemba effect with two disconnected initial-temperature windows. In this sense, quantization fundamentally reshapes the anomalous-relaxation landscape that the system can access.

The origin of this difference lies in a change of the relaxation bottleneck. Classically, relaxation is controlled by thermal barrier crossing \cite{kumar2020exponentially, Kumar/etal:2022, malhotra2024double, lu2017nonequilibrium, Hanggi_RMP} and hence by the height $V_0$ of the barrier potential $V(x)$, see the illustration in Fig.~\ref{fig:1}. In the quantum case, by contrast, relaxation proceeds through transitions among discrete eigenstates \cite{breuer2002theory, Rivas/Huelga:2012, Carmichael:1993, hayakawa2026quantumtunnelingmpembaeffect}, with the relevant scale set by low-lying weakly coupled levels whose splittings lie far below the barrier height (see Fig.~\ref{fig:1}). Anomalous cooling channels can therefore remain effective even where classical over-barrier motion is strongly suppressed. Taken together, these results provide a direct classical-to-quantum comparison in the paradigmatic Mpemba platform, identifying quantization itself as a route to anomalous relaxation in otherwise classically inaccessible regimes. These results point to ultracold atoms in tailored optical traps as a natural setting for experimental exploration \cite{bloch_quantum_2012}. Ultracold atoms in double-well potentials have been used to probe and illustrate quantum properties since the realization of quantum gases \cite{Andrews97}. Since then, tight control over external potentials have enabled a plethora of studies, including investigating the dynamics in Josephson-contact-type potentials of interacting Bose-Einstein condensates \cite{Albiez2005} or ultracold Fermions \cite{Luick2020}, or the entanglement formation and interference of single atoms in quantized degrees of freedom \cite{Kaufman2014, Murmann2015}.

An important difference to classical systems is the strong isolation of ultracold quantum gases, and relaxation usually is absent on experimentally relevant time scales \cite{Bloch_2008,ueda_quantum_2020}. 
In order to intentionally induce and control relaxation, two approaches have been used. First, the quantum gas is brought into an interacting state, and this interaction can lead to relaxation and thermalization \cite{rigol_2008,ueda_quantum_2020}. Second, the quantum system can be brought into contact with another ultracold bath that induces thermalization \cite{Spethmann_2012,Scelle_2013,Catani_2012}.

We note that, in the quantum regime, the roles of the two minima of a bistable potential can be reversed, as illustrated in Fig.~\ref{fig:1}. Specifically, in an asymmetric potential whose secondary minimum is shallower than the primary one, the quantum ground state may become predominantly localized in the shallower minimum. By contrast, the classical equilibrium state is determined by the Boltzmann statistics and the ground state therefore remains associated with the deeper minimum. 
Although such a quantum-induced inversion of the ground state location is not required for the emergence of the quantum Mpemba effect, it substantially enhances its magnitude, as shown in the Supplemental Material (SM). In the following, we focus on a potential exhibiting this inversion, as shown in Fig.~\ref{fig:1}.

\begin{figure}[htp!]
    \centering \includegraphics[width=0.95\columnwidth]{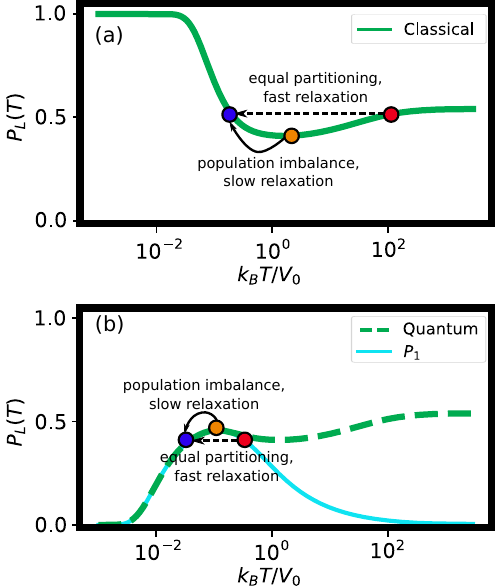} 
    \caption{Key ingredient for the Mpemba effect: non-monotonic temperature dependence of the thermal probability $P_L$ to occupy a (left) domain. Both (a) classical (solid green) and (b) quantum (dashed green) systems exhibit such behavior, with the black dashed horizontal arrows highlighting pairs of temperatures with identical domain occupation probabilities (equal partitioning), associated with rapid relaxation from hotter system (marked by red circle) to a colder one (blue circle). When the domain occupations associated with the initial and target temperatures do not match (as e.g. for the orange and blue circles), relaxation must overcome a bottleneck associated with redistributing the population between the domains and therefore takes longer despite starting close to the target temperature, thereby manifesting the Mpemba effect. In the quantum system (b), occupation of the left domain in the low-temperature regime is entirely determined by the population $P_1 = \bra 1\hat\rho_T\ket1$ of the state $\ket1$ (cyan solid), since $\ket1$ is the only state strongly localized in the left domain (see Fig.~\ref{fig:1}); higher-energy delocalized states are negligibly populated and therefore do not contribute.}
    \label{fig:2}
\end{figure}

\begin{figure*}[ht]
    \centering \includegraphics[width=0.9\linewidth]{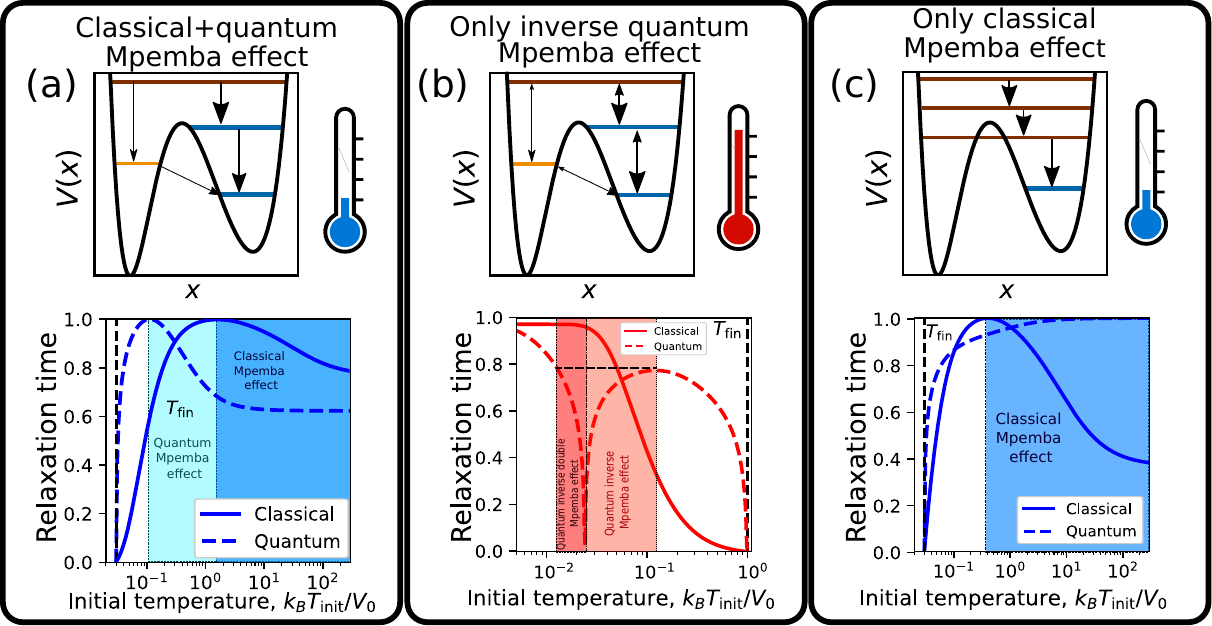} 
    \caption{(a) A Mpemba effect (reduction of the relaxation time upon increasing the initial temperature) for the cooling process is observed in both classical and quantum systems, where the latter emerges at a much colder temperature range compared to the classical counterpart. The temperature ranges over which the relaxation time decreases are indicated by rectangles (blue for the classical system, extended with a cyan region for the quantum system). (b) In the classical case, the inverse Mpemba effect (reduction of the relaxation time upon decreasing the initial temperature) is suppressed when the system is heated only up to a target temperature below the barrier height; however, a quantum-induced inverse Mpemba effect still persists. The relaxation time exhibits two extrema: one upon its drop to zero and another marked by the horizontal dashed line. For the light-red rectangle, the relaxation time decreases with decreasing initial temperature, corresponding to the inverse Mpemba effect; for the red rectangle, two distinct hotter initial temperatures yield identical relaxation times, making the effect double.
    (c) For a potential without a weakly coupled eigenstate the quantum Mpemba effect is suppressed, with the temperature range in which the classical Mpemba effect occurs marked by a blue rectangle. In all panels, the relaxation time is normalized with respect to one.}
    \label{fig:3}
\end{figure*}

To observe the Mpemba effect, in both classical and quantum systems we start from the system in the thermal state at the initial temperature $T = T_{\rm init}$, which is at $t=0$ abruptly changed to the target temperature $T = T_{\rm fin}$. In the classical case, the thermal probability density yields the Boltzmann distribution $P(x; T) \sim e^{-\beta V(x)},\, \beta = 1/k_B T$; while in the quantum case, the density matrix $\hat\rho$ of a thermal state reads
\begin{subequations}
\begin{equation}
    \hat \rho_{T} = \frac{e^{-\beta \hat{H}}}{\textrm{Tr}\left(e^{-\beta \hat{H}}\right)},
\end{equation}
where $\hat H$ is the quantized Hamiltonian,
\begin{equation}
    \hat{H} = \frac{\hat{p}^2}{2m}+\hat V(\hat x)
    \label{eq:ham}
\end{equation}
with momentum operator $\hat{p}$, position operator $\hat x$, potential energy $\hat V(\hat x)$, and particle mass $m$.
\end{subequations}

For the considered system with two well-separated domains, the relaxation process to the steady-state occurs on two different timescales. The probability distribution within each domain relaxes rapidly towards a local equilibrium, while the re-distribution between the left and right domains is much slower. In the classical case, the transition between the domains are rare because they require thermally activated barrier crossing, while in the quantum case those are suppressed by the weak coupling between the states associated with the two domains.
Crucially, the domain occupation probabilities 
$P_{L,R}(T)$ depends on the temperature; since $P_R(T) = 1 - P_L(T)$, it is sufficient to consider the temperature dependence of the left domain $P_L(T)$. If the function $P_{L}(T)$ is non-monotonic,
two temperatures $T^{(0)}_{\rm init}$ and $T^{(0)}_{\rm fin}$ can be found which have the {\it same\/} probability, i.e. 
$P_{L}(T^{(0)}_{\rm init}) = P_{L}(T^{(0)}_{\rm fin})$, see the illustration in Fig.~\ref{fig:2}. Then the initial thermal state relaxes quickly towards the target one, since no transition between the domains is needed. However, for temperatures different from $T^{(0)}_{\rm init}$ the transition over the bottleneck is required and the relaxation time is longer. Hence, a non-monotonic behavior in $P_{L}(T)$ naturally induces a Mpemba effect. 

Classically, since the relevant temperature scales are naturally separated by the barrier height $V_0$, 
the left-domain occupation probability behaves as
\begin{align}
    P_L(T) \approx \begin{cases}
        \frac{1}{1+e^{-\frac{\Delta V}{k_B T}}}, & k_B T\lesssim V_0,\\
        \frac{\Delta x_L}{\Delta x_L+\Delta x_R}, & k_B T \gg V_0,
    \end{cases}
\end{align}
where $\Delta x_{L,R}$ denote the spatial extents of the left and right domains, respectively, arising from the finite confinement of the experimental setup, and $\Delta V$ is the energy difference between the corresponding potential minima. In this picture, $P_L(T)$ initially follows the Boltzmann weighting and increases monotonically with temperature. However, in the high-temperature regime, the nearly homogeneous distribution over the finite domain causes the occupation of the domains to become dominated by geometric confinement, leading $P_L(T)$ to approach $\Delta x_L/(\Delta x_L+\Delta x_R)$. As a result, the probability exhibits a non-monotonic dependence on temperature, with its extremum located above $V_0$. For more details on the confinement-induced classical Mpemba effect, we refer the reader to \cite{Liu/Hayakawa:2026, Liu/etal:2026}.

For the quantum Mpemba effect to emerge, multiple slow modes need to exist in the system for the metastability of one of the states \cite{Macieszczak/etal:2021,Macieszczak/etal:2016}. Without loss of generality, we assume it to be $\ket 1$, implying that this state is weakly coupled to the rest of the system, as shown in Fig.~\ref{fig:1}. The long-time relaxation is then governed by the slowest decay mode ($\ket 1 \to \ket 0$ for the cooling process) and therefore depends on the probability $P_1$ that the system initially occupies the state associated with this transition channel,
\begin{equation}
    P_1(T) = \frac{e^{-\beta \epsilon_1}}{\sum_n e^{-\beta \epsilon_n}} = \frac{e^{-\beta \Delta \epsilon}}{1 + \sum_{n\ge 1}e^{-\beta (\epsilon_n - \epsilon_0)}},
\end{equation}
where $\epsilon_n$ are the eigen-energies of the Hamiltonian \eqref{eq:ham}, and $\Delta \epsilon = \epsilon_1-\epsilon_0$. This population exhibits non-monotonicity naturally as it approaches zero for $T=0$ and $T \to \infty$ temperatures, with an extremum occurring at $k_BT \sim \Delta \epsilon < V_0$. Hence, the quantum-enabled Mpemba effect typically emerges at lower temperatures, making it accessible in regimes where it is not observable in the classical case. Analytical considerations are provided in the End Matter.

In Figure~\ref{fig:3} we outline all possible scenarios for the direct comparison between classical and quantum systems. When the effect emerges in both classical and quantum systems upon cooling $T_{\rm init}>T_{\rm fin}$ (Fig.~\ref{fig:3}(a)), it occurs over entirely different temperature ranges, with the quantum Mpemba effect emerging at much lower initial temperatures. 

For heating $T_{\rm init}<T_{\rm fin}$, an \textit{inverse} Mpemba effect is observed when the initially cold system heats up faster than the initially hotter one. Moreover, the relaxation time not just non-monotonic in $T_{\rm init}$ but also exhibits \textit{two} extrema (as shown in Fig.~\ref{fig:3}(b)). While the analogous classical system is referred to as a \textit{double} Mpemba effect. Previously, it has been theoretically predicted for a classical system upon cooling~\cite{malhotra2024double}; here we find a novel scenario in a quantum system, where the inverse Mpemba effect emerges under a heating protocol.

For the target temperature set below the height of the energy barrier $k_B T_{\rm fin} \ll V_0$ (Fig.~\ref{fig:3}(b)), the Mpemba effect is therefore suppressed classically but still persists in the quantum system. Finally, as the quantum Mpemba effect requires a weakly-coupled state to be present in a system, for a potential with all eigenstates strongly coupled to their nearest neighbors (Fig.~\ref{fig:3}(c)), the quantum Mpemba effect is suppressed but still occurs classically. 

In order to assess the realization in standard ultracold atom experiments, we may consider typical experimental values, see for example Refs.~\cite{Shin:2004,Albiez2005,petrucciani_2026}. Typically, a Bose-EInstein condensate of $10^3 \ldots 10^4$ atoms are trapped in far-detuned optical double-well potentials with transverse trapping frequencies of the order $1\,$kHz. For optical tweezer potentials, this frequency can be significantly increased to some $10\,$kHz. Typical temperatures of the gas are $k_B T \approx 10\ldots 1000\,$nK, resulting in ratios $k_BT/(\hbar \omega) \approx 0.1\ldots 100$. This suggests that the characteristic temperature range $k_BT/(\hbar \omega) \approx1.4$ (see the SM) used in our theoretical proposal is experimentally feasible.

In this work, we have demonstrated a quantum-enabled Mpemba effect that generally emerges at significantly lower temperatures than its classical counterpart for the same underlying Hamiltonian. Since the Mpemba effect can occur in both cooling and heating processes, our results point to a broad range of potential strategies for efficient temperature control in quantum systems, making them particularly relevant for future experimental investigations. In particular, the proposed model offers a promising platform for the rapid preparation of initial states in systems of ultra-cold atoms confined in optical traps \cite{Savard/etal:1997, Gross/Bloch:2017, Browaeys/Lahaye:2020, Schaefer/etal:2020, Holman/etal:2026}. Such fast state preparation is often a key requirement in quantum technologies, for example in strongly correlated molecular systems \cite{Berry/etal:2025}, quantum computing architectures \cite{Fomichev/etal:2024}, and the preparation of many-body ground states \cite{Zhan/etal:2026}. Another central result of the present work is the observation of a double inverse quantum Mpemba effect. 
While double Mpemba effects have barely been discussed in the existing literature and exclusively for the classical setups \cite{malhotra2024double}, our results suggest a promising platform for studying such complex Mpemba phenomena in quantum systems. Most importantly, the double inverse effect reported here is driven by genuinely quantum mechanisms and emerges in entirely new parameter regimes absent in the classical counterpart. Our findings open new directions for exploiting anomalous relaxation phenomena in nonequilibrium quantum systems, including the recently emerged field of quantum active matter \cite{Adachi/etal:2022, Khasseh/etal:2023, Yamagashi/etal:2024, Takasan/etal:2024, Antonov/etal:2025, Nadolny/etal:2025, Penner/etal:2025} -- a class of systems obeying quantum statistics while continuously consuming and converting energy into persistent driven motion -- which may be regarded as a paradigmatic setting for nonequilibrium quantum dynamics.

\section*{End Matter}

\textit{Derivation of the relaxation dynamics}
-- In both classical and quantum systems, we consider the Liouvillian dynamics, with
\begin{subequations}
    \begin{equation}
        \partial_t P = {L} P
    \end{equation}
for the classical system, and 
\begin{equation}
    \partial_t \hat\rho = \mathcal{L}\hat \rho
\end{equation}
\end{subequations}
for the quantum one. Here $\hat L$ is the classical Liouvillian operator and $\mathcal{L}$ is the quantum superoperator. In both cases we consider dissipative dynamics for which the thermal state is a stationary state of the evolution. For the classical case, we consider the overdamped dynamics with external potential $V(x)$, for which the Liouvillian is a Fokker-Planck equation of the form:
\begin{equation}
     L = \partial_x\left(\frac{V'(x)}{\gamma} + D\partial_x\right),
\end{equation}
where $\gamma$ is the damping rate and $D$ is the diffusion coefficient. 
For the quantum system, we consider dissipative Lindblad dynamics \cite{Lindblad:1976}, for which the Liouvillian reads (see also the SM):
\begin{align}
    & {\mathcal{L}}\hat X=-\frac{i}{\hbar}\left[\hat{H}, \hat X\right] \nonumber\\
    &+ \sum_{\omega} \gamma(\omega)\left(\hat{A}(\omega) \hat X\hat{A}^\dagger(\omega) - \frac{1}{2}\left\{ \hat{A}^\dagger(\omega)\hat{A}(\omega), \hat X \right\} \right),
\label{eq:Lindblad:0}
\end{align}
where $\hat X$ is a generic operator, $\omega$ labels the decay channels, $\hat A(\omega)$ is the transition operator,
\begin{equation}
    \hat{A}(\omega) = \sum\limits_{\epsilon_n-\epsilon_m = \hbar\omega} \ket n \bra n \hat{A}\ket m\bra m,
\end{equation}
where the indices $m,n$ label energy eigenstates, while $\hbar\omega = \epsilon_n-\epsilon_m$ specifies the energy difference associated with the transition $\ket m \to \ket n$; and $\gamma(\omega)>0$ is the corresponding transition rate. In this paper, we consider $\hat A = \hat x$, corresponding to the dipole coupling.

The evolution of classical probability reads \cite{Risken}:
\begin{equation}
    P(x,t) = P(x; T_{\rm fin}) + \sum_{n=1}^{\infty} c_n \phi_n(x)e^{\lambda_n t},
\end{equation}
where $\phi_n(x), \lambda_n$ are the eigenfunctions and eigenvalues of the Liouvillian operator, $0 = \lambda_0 > \lambda_1 > \ldots$ and $c_n$ set the initial conditions. For a considered bistable potential with a metastable state, the relaxation is governed by the slowest decay mode:
\begin{equation}
    P(x,t) \approx P(x; T_{\rm fin}) + c_1\phi_1(x)e^{-|\lambda_1| t},
    \label{eq:FPE-SDM}
\end{equation}
The distance measure to the steady-state is defined as
\begin{equation}
    \Delta(t) = \int_{-\infty}^{\infty} dx|P(x,t) - P(x;T_{\rm fin})|,
\end{equation}
and follows the exponential relaxation law:
\begin{align}
    \Delta(t)\approx e^{-|\lambda_1| t} \int_{-\infty}^{\infty} dx |c_1\phi_1(x)|.
    \label{eq:classical-decay}
\end{align}
Generally, the relaxation time for the exponential law $\Delta(t)= a(T)e^{-|\lambda|_1t}$ is given by
\begin{equation}
    t_{\rm relax} = \frac{\ln\!\left(a(T)/\varepsilon_0\right)}{|\lambda_1|},
\end{equation}
where $\varepsilon_0 = \Delta(t_{\rm relax})$ is the tolerance distance chosen to be arbitrary small. Here, it is evident that the emergence of the Mpemba effect requires a non-monotonic temperature dependence of the prefactor $a(T)$.
For a considered bistable potential with a metastable state, we can attribute the slowest decay mode with the population imbalance of the domains between the initial and final states. The direct integration of \eqref{eq:FPE-SDM} at $t=0$ yields:
\begin{equation}
    \int_{-\infty}^{\infty} dx|c_1 \phi_1(x)| \approx 2 |P_L(T_{\rm init}) - P_L(T_{\rm fin})|,
\end{equation}
where we have assumed that all faster relaxation modes are suppressed. In this limit, the thermal state is well approximated by localization in the two potential wells,
\begin{equation}
P(x; T) \approx P_L(T)\delta(x - x_a) + P_R(T)\delta(x - x_b),
\end{equation}
where $x_{a,b}$ are positions of the minima in the left and right domains, respectively.
Consequently, equation \eqref{eq:classical-decay} can be written as
\begin{equation}
    \Delta(t) \approx 2e^{-|\lambda_1| t}|P_L(T_{\rm init}) - P_L(T_{\rm fin})|,
\end{equation}
where the exponential prefactor accounts for the initial mispopulation of the left domain discussed in the main text.

\begin{figure}[htp!]
\vspace{2ex}
    \centering \includegraphics[width=0.7\linewidth]{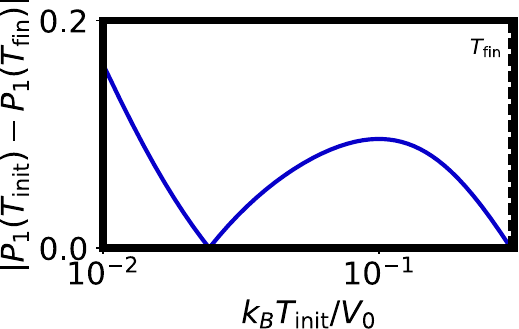} 
    \caption{The population imbalance $|P_1(T_{\rm init}) - P_1(T_{\rm fin})|$ of the first excited state for the system in Fig.~\ref{fig:3}(b).}
    \label{fig:4}
\end{figure}

For the quantum system, the density matrix evolution can be cast in a relatively transparent form by using the spectral decomposition of the Lindblad operator \eqref{eq:Lindblad:0}. In what follows we denote by $\hat\Phi_{\Lambda}$ the right eigenvectors at eigenvalue $\Lambda$, satisfying the eigenvalue equation $\mathcal L\hat\Phi_\Lambda= \Lambda\hat\Phi_\Lambda$. The corresponding left eigenvectors read $\hat L_\Lambda\mathcal L=\Lambda \hat L_\Lambda$ and the orthogonality relation holds ${\rm Tr}(\hat L_{\Lambda'}\hat\Phi_\Lambda)=\delta_{\Lambda',\Lambda}$. The eigevectors form a biorthogonal basis in the absence of exceptional points. In this case the time evolution of the density operator can be cast into the form:
\begin{equation}
    \hat\rho(t) = \hat{\rho}_{T_{\rm fin}} + \sum_{n=1}^{\infty} c_n \hat\Phi_ne^{\Lambda_n t},
\end{equation}
where $c_n={\rm Tr}(\hat L_n\hat\rho(0))$ is the overlap between the initial density matrix and the eigenvector at eigenvalue $\Lambda_n$ and the numbering between eigenvalues is according to the convention $0=\Lambda_0 > \textrm{Re}(\Lambda_1) > \ldots$.
The long-time relaxation is then governed by the slowest decay mode ($\ket 1 \to \ket 0$ for the cooling process) and therefore depends on the probability that the system initially occupies the state associated with this transition channel. This information is encoded in the coefficient $c_1$ \cite{vanKampen2007}
\begin{equation}
    c_1 = \textrm{Tr}(L_1 \hat\rho_{t =0})\,.
\end{equation}
Since the initial thermal-state density matrix $\hat\rho_{t=0}$ is diagonal in the energy basis, 
then
\begin{equation}
    c_1 = \sum_{n=0}\alpha_n P_n(T_{\rm init}).
\end{equation}
 where $\alpha_n=\langle n|L_1|n\rangle$,
with the dominant contribution arising from $\alpha_1$ as the metastable state $\ket 1$ is well localized in the left domain and weakly coupled to the other states, in agreement with numerical results.
Within the two-level approximation involving only $\ket 0$ and $\ket 1$, this expression reduces to
\begin{equation}
    c_1 = (\alpha_1 - \alpha_0)P_1(T_{\rm init}) + \alpha_0.
\end{equation}
Given that $\alpha_0$ does not depend on the initial temperature and $c_1=0$ for $T_{\rm init} = T_{\rm fin}$, we obtain
\begin{equation}
    c_1 = (\alpha_1 - \alpha_0)\left[P_1(T_{\rm init}) - P_1(T_{\rm fin})\right].
\end{equation}
Upon cooling the quantum system to a target temperature such that the ground state is predominantly occupied $P_0(T_{\rm fin}) \to 1$, the slow Liouvillian eigenoperator simplifies to
\begin{equation}
    \hat\Phi_1 \approx \ket 1 \bra 1 - \ket 0 \bra 0.
\end{equation}
For the trace distance defined as $\|A\|_{1}\equiv \operatorname {Tr} [{\sqrt {A^{\dagger }A}}]$ \cite{Nielsen},
\begin{equation}
    \mathcal{D}(t) = \frac{1}{2}||\hat \rho(t) - \hat{\rho}_{T_{\rm fin}}||_1,
\end{equation}
and using the single-mode approximation, one obtains:
\begin{align}
    \mathcal{D}(t) &\sim \frac{1}{2} |P_1(T_{\rm init}) - P_1(T_{\rm fin})| \big|\big|\ket 1\bra 1 - \ket 0\bra 0\big|\big|_1e^{-|\textrm{Re}(\Lambda_1)| t} \nonumber\\
    &= |P_1(T_{\rm init}) - P_1(T_{\rm fin})|e^{-|\textrm{Re}(\Lambda_1)| t},
    \label{eq:quantum-relax}
\end{align}
where a non-monotonic temperature dependence of $P_1(T)$ gives rise to the quantum Mpemba effect, as discussed in the main text. 
For the heating process, the prefactor $|P_1(T_{\rm init}) - P_1(T_{\rm fin})|$ of this law exhibits two extrema
(see Fig.~\ref{fig:4}), thereby giving rise to the double effect discussed in the main text.

To achieve a controlled comparison between classical and quantum systems, we consider a quantum Lindblad dynamics defined on the same underlying energy landscape as the classical system and coupled to a thermal reservoir leading to a thermal stationary state. For the considered dissipative dynamics, which relaxes to thermal equilibrium, the emergence of the Mpemba effects is governed by the spectral properties of the Liouvillian and the presence of metastable states
rather than by microscopic details of the Liouvillian.
Therefore, it does not matter here that the form of the dissipative dynamics generally differs between classical colloidal systems and quantum systems \cite{antonov2025modeling}. In the SM, we outline parameters of the systems for the scenarios described in Fig.~\ref{fig:3}.

\end{document}